\documentclass[%
 aip,
 amsmath,amssymb,
preprint,%
]{revtex4-1}

\usepackage{graphicx}% Include figure files
\usepackage{dcolumn}% Align table columns on decimal point
\usepackage{bm}% bold math
\usepackage{braket}
\usepackage[utf8]{inputenc}
\usepackage[T1]{fontenc}
\usepackage{mathptmx}
\usepackage{color}
\usepackage{threeparttable}

\begin{document}

\preprint{}

\title[]{Theory of Frequency Fluctuation of Intramolecular Vibration in Solution Phase: Application to C--N Stretching Mode of Organic Compounds}
% Force line breaks with \\

\author{Naoki Negishi}
 \email{negishi-naoki@g.ecc.u-tokyo.ac.jp }
\author{Daisuke Yokogawa$^{*}$}%
 \email{c-d.yokogawa@g.ecc.u-tokyo.ac.jp }
\affiliation{ 
Department of Multidisciplinary Science, Graduate School of Arts and Sciences, The University of Tokyo, 
Komaba, Meguro-ku, Tokyo 153-8902, Japan}%

\date{\today}% It is always \today, today,

\begin{abstract}
We formulate frequency fluctuations of intramolecular vibrations of a solute by exploring the fluctuation of the electrostatic potential by solvents. We present a numerical methodology for estimating the frequency fluctuations; the methodology is based on the reference interaction site model self-consistent field with constrained spatial electron density distribution, a the theoretical model of solvation fields based on classical statistic mechanics. By applying the present theory to the C--N stretching vibrations of several nitrile compounds, our estimated frequency fluctuation scale and bandwidth shift by changing solvent kinds reproduced the experimental data.
Further, we regard the standard deviation of the electrostatic potential as the multiple random variables for analyzing the frequency fluctuations. Our results reveal that the dominant fluctuation of the electrostatic field is almost parallel to the vibrational axis. Additionally, the fluctuations of electrostatic potential become spatially nonuniform as the solvents form stronger hydrogen bonds with the solute. The development of the solvation field confirms that the nonuniformity of the electrostatic field is crucial to the frequency fluctuation.
\end{abstract}

\maketitle

\section{\label{sec:level1}INTRODUCTION}
Spectroscopy of far infrared (IR) wavelength region provides rich information about the structures and inner motions in many materials. The spectral lines of the IR absorption spectra of molecules display multiple peaks that corresponds to the eigen frequencies of their intramolecular vibration. Numerous IR spectra data revealed the frequency regions of various molecular vibrations. The overall rotational motion of gaseous-phase molecules can aid the assignment of all peaks to intramolecular motion if the energy resolution exceeds the gap between rotational energy levels.\cite{PhysRev.56.1113,PhysRevLett.97.193001,PhysRevLett.120.263002,OLMAN196462}\par
Conversely, the solution-phases case is more complicated. Compared with the gaseous phase, peak broadening sometimes produces narrower or broader. Generally, the narrower case proceeds from the sufficiently rapid relaxation of the rotational motion\cite{rot_relax} ($1\sim 10$ ps) compared with the rotational frequency of the molecules ($> 10$ ps) owing to the scattering between the solute and solvent molecules. In the broader case, the broadening mechanism is much more complicated and yields various peak structures. One of the simplest explain this is that the friction to the oscillator\cite{Drude} causes the dephasing of vibrational motion. This represents the physical interpretation, as the dephasing rate is determined by the viscosity of the solvent surroundings. Otherwise, the energy fluctuation of solvation energy\cite{DebyeHuckel1923,Marcus1964,BagchiChandra1991} can also account for the peak broadening. Particularly, the energy scale of the electrostatic interaction between a solute and solvent (1$\sim$10 kcal/mol) is nearly equal to the intramolecular vibration. Therefore, a strong electrostatic coupling such as hydrogen bonding induces large fluctuations in the intramolecular vibrational frequencies. The extant IR spectroscopic studies of the C--N stretching mode of nitrile compounds in various solvents\cite{EXPT1,EXPT2,EXPT3,10.1063/5.0082969,Ion_CN,ACN_MeSCN_EXPT4} revealed the solvatochromic shift in the peak position and the peak-broadening tendencies.\par
Notably, the fluctuation of the solvation structure must be quantified if peak broadening mainly proceeds from the fluctuation of the electrostatic interaction between molecules. Although molecular dynamics (MD) is the standard method for directly computing the solute--solvent mixed phase, it is accompanied by bottlenecks such as the computational cost of simulating Raman scattering as the time correlation of the system's polarizability must be investigated. Additionally, classical MD cannot accurately determine the dielectric relaxation caused by the solvent reorganization. This can be determined by multiscale quantum simulations, such as quantum mechanics / molecular mechanics (QM/MM)\cite{QMMM1,QMMM2,QMMM3} or ultimately {\it ab initio} MD\cite{AIMD}, both of which will be much more computationally demanding.\par
Thus, in this study, we developed a framework based on the reference interaction site model (RISM)\cite{Hirata1981}, which is a statistical mechanic-based integral equation theory for molecular liquid. Notably, the RISM theory has been, successfully combined with the molecular orbital theory to form an extended framework known as the RISM-self consistent field\cite{Tenno1994,TENNO1993391,Sato1996} with constrained spatial electron distribution (RISM-SCF-cSED)\cite{cSED,SEDD,SEDD-new}, which simulates solution systems with chemical accuracies. We recently developed the RISM-SCF-cSED to compute the static fluctuation of the solvation field and quantified the broadening bandwidth of the electronic absorption spectral lines of the solution.\cite{RISM-DMRG,NegishiYokogawa2023,NegishiYokogawa2024} Thus, by applying our RISM-SCF-cSED framework, we present a framework for evaluating the fluctuation of the intramolecular vibrational frequency in solvent surroundings.\par
The remainder of this paper is organized as follows: Section \ref{sec:level2} presents an overview of the theoretical framework for evaluating the frequency fluctuation of a solute molecule by first defining the potential energy surface (PES) based on the RISM-SCF-cSED framework, followed by exploring the electrostatic potential as the multiple random variables. The computational details are summarized in Section\ref{sec:level3}, and Section\ref{sec:level4} reports the investigation of the frequency fluctuation of the C--N stretching of nitrile molecules in various polar solvents. Thus, the deployed framework reproduced the experimental energy scale and solvatochromism of bandwidth. Section\ref{sec:level5} presents the summary and conclusion of the study.\par
\section{\label{sec:level2}THEORY}
\subsection{Fluctuation of vibrational frequency based on RISM}
First, we derive the fluctuation of the vibrational frequencies of a solute molecule. In this study, we introduce the fluctuation of PES using the RISM-SCF-cSED method. In the RISM-SCF-cSED framework, the PES $\mathcal{A}({\bf R})$ is defined as
\begin{eqnarray}
    \mathcal{A}({\bf R})=\braket{\Psi({\bf R})|\hat{H}_{\rm e}({\bf R})|\Psi({\bf R})}+\mu({\bf R}),
\end{eqnarray}
where $\Psi({\bf R})$, $\hat{H}_{\rm e}({\bf R})$, and $\mu({\bf R})$ are the electronic ground state, the electronic Hamiltonian of the solute, and excess chemical potential under the atomic configuration of the solute given by ${\bf R}$. The excess chemical potential under the approximation of the hyper-netted chain (HNC) closure\cite{HNC} is given by the following:
\begin{eqnarray}
    \mu({\bf R})=k_BT\sum_{s,\alpha}\rho_s\int d{\bf r} \frac{1}{2}h_{\alpha s}^2(r;{\bf R})-\frac{1}{2}h_{\alpha s}(r;{\bf R})c_{\alpha s}(r;{\bf R})-c_{\alpha s}(r;{\bf R}),
\end{eqnarray}
where $h_{\alpha s}(r;{\bf R})$ and $c_{\alpha s}(r;{\bf R})$ are total and direct correlation functions between the atomic sites of the solute $\alpha$ and solvent $s$, respectively. $k_B$ and $T$ are the Boltzmann constant and temperature, respectively. Both correlation functions are determined by using the RISM equation and HNC closure relation in Ref\cite{HNC}. The RISM theory is based on the integral equation theory with the microscopic model potential to obtain the solvation structure. The interaction between the atomic sites of the solutes $\gamma$ and solvents $s$,  $u_{\gamma s}(r)$, is given, as follows:
\begin{eqnarray}
   u_{\gamma s}(r)&=&u_{\gamma s}^{\rm nES}(r)-q_s\int d{\bf r'}\frac{\tilde{\rho}_{\gamma}({\bf r'})}{|{\bf r'}-{\bf r}|}+\frac{q_sZ_{\gamma}}{r}\label{eq:potential_UV}.
\end{eqnarray}
Here, $u_{\gamma s}^{\rm nES}$ is the non-electrostatic site--site pair potential, $Z_\gamma$ is the nuclear charge of the solute site $\gamma$, $q_s$ is the atomic charge of the solvent site $s$, and $\tilde{\rho}_{\gamma}$ is the electron density distribution around the atomic site of the solute site $\gamma$. Here, we assume that $\tilde{\rho}_{\gamma}$ can be expanded by auxiliary basis sets ${\bf f}$ and the fitting coefficients ${\bf d}$\cite{cSED,SEDD,SEDD-new} as follows:
\begin{eqnarray}
    \tilde{\rho}_{\gamma}({\bf r})=\sum_{\zeta\in\gamma}d_{\zeta}f_{\zeta}({\bf r}).\label{eq:fitting}
\end{eqnarray}
We determine ${\bf d}$ to optimize the electron density obtained from the {\it ab initio} calculation following Ref.\cite{cSED}\par
To determine the solute and solvent structures, we minimize the free energy $\mathcal{A}({\bf R})$. Further, we derive the variational equations for $\Psi$, ${\bf h}$, and ${\bf c}$ as well as the energy gradient to determine the equilibrium geometry of ${\bf R}$. The variational equations for ${\bf h}$ and ${\bf c}$ correspond to the RISM equation and HNC closure relation. The variation of $\Psi({\bf R})$ yields the following: 
\begin{eqnarray}
    \Braket{\delta\Psi({\bf R})|\hat{H}_{e}({\bf R})+^t{\bf V}({\bf R})\hat{\bf Q}({\bf d};{\bf R})-\hat{\Lambda}|\Psi({\bf R})}=0.\label{eq:gen_VP}
\end{eqnarray}
Similar to  the form of the Schr\"{o}dinger equation, $\hat{\Lambda}$ denotes the Lagrange multiplier for satisfying the orthonormality of the wavefunction. The column vector ${\bf V}$ is derived from the solvent structure as follows:
\begin{eqnarray}
    V_{\zeta}&=& \frac{\partial \Delta\mu}{\partial d_{\zeta}}\notag\\
    &=&-\sum_{s}k_BT\rho_s q_s\int d^3{\bf r}\int d^3{\bf r'}h_{\gamma s}(r)\frac{f_{\zeta}({\bf r}')}{|{\bf r}-{\bf r}'|}.\,\,\, {\rm for} (\zeta \in \gamma)
\end{eqnarray}
 The second term in Eq.(\ref{eq:gen_VP}) corresponds to the electrostatic interaction between a solute and solvents determined by the total correlation function ${\bf h}$. Therefore, Eq.(\ref{eq:gen_VP}) determines the wavefunction $\Psi$ when ${\bf h}$ is determined.\par
We demonstrate that $\mathcal{A}({\bf R})$ can be defined when the solvent structures, ${\bf h}({\bf R})$ and ${\bf c}({\bf R})$, and the electronic state, $\Psi({\bf R})$, are determined. Next, we extend $\mathcal{A}({\bf R})$, which includes the degree of freedom for the deviation of fluctuation of the solute--solvent electrostatic interaction. Here, ${\bf V}({\bf R})$ represents the key variables. We define the deviations of ${\bf V}({\bf R})$ as $\Delta {\bf V}$ due to the thermal fluctuations of the solvents. When ${\bf V}({\bf R})$ is deviated toward ${\bf V}({\bf R})+\Delta {\bf V}$, how should $\mathcal{A}({\bf R})$ be deviated? Let us expand the degree of freedom in the PES as $\mathcal{A}({\bf R},\Delta{\bf V})$ and approximate $\mathcal{A}$ within the second order of $\Delta {\bf R}={\bf R}-{\bf R}_0$ and first order of $\Delta{\bf V}$ as follows:
\begin{eqnarray}
    \mathcal{A}({\bf R},\Delta{\bf V})&\sim& \mathcal{A}({\bf R}_0,0)+\sum_{a,b}\frac{1}{2}\left(\frac{\partial^2 \mathcal{A}}{\partial R_a\partial R_b}\right)_{{\bf R}={\bf R}_0,\Delta{\bf V}=0} \Delta R_a\Delta R_b\notag\\
    &+&\sum_{\zeta}\left(\frac{\partial\mathcal{A}}{\partial V_{\zeta}}\right)_{{\bf R}={\bf R}_0,\Delta{\bf V}=0}\Delta V_{\zeta}+\sum_{\zeta}\sum_{a}\left[\frac{\partial}{\partial V_{\zeta}}\left(\frac{\partial \mathcal{A}}{\partial R_a}\right)_{{\bf R}={\bf R}_0}\right]_{\Delta{\bf V}=0}\Delta R_a\Delta V_{\zeta}\notag\\
    &+&\sum_{\zeta}\sum_{a,b}\frac{1}{2}\left[\frac{\partial}{\partial V_{\zeta}}\left(\frac{\partial^2 \mathcal{A}}{\partial R_a\partial R_b}\right)_{{\bf R}={\bf R}_0}\right]_{\Delta{\bf V}=0}\Delta R_a\Delta R_b\Delta V_{\zeta}\notag\\
    &:=&\mathcal{A}({\bf R}_0)+\sum_{a,b}\frac{1}{2}K_{ab}^{(0)}\Delta R_a\Delta R_b\notag\\
    &+&\sum_{\zeta}A_{\zeta}^{(1)}\Delta V_\zeta+\sum_{\zeta}\sum_{a}G_{a,\zeta}^{(1)}\Delta R_a\Delta V_\zeta+\sum_{a,b}\sum_{\zeta}\frac{1}{2}K_{ab,\zeta}^{(1)}\Delta R_a\Delta R_b\Delta V_{\zeta},
    \label{eq:taylor_VR}
\end{eqnarray}
where ${\bf R}={\bf R}_0$ denotes the equilibrium geometry of the solute. The first and second terms in Eq.(\ref{eq:taylor_VR}) represent the thermal average of $\mathcal{A}({\bf R},\Delta{\bf V})$, as obtained using the RISM-SCF Hessian method.\cite{HESS_RISM} Additionally, the residual third and fourth terms represent the first-order deviation of the thermal fluctuation of ${\bf V}({\bf R})$. The third and fourth terms denote the displacement of the potential energy level and equilibrium geometry irrelevant to the frequency fluctuation of the intramolecular vibration. Conversely, the fifth term represents the fluctuation of the curvature of $\mathcal{A}({\bf R})$; thus, it is directly relevant to the frequency fluctuation.\par
Furthermore, $\mathcal{A}({\bf R})$ is a fundamentally assumed relaxed approximation, indicating the adiabatic relaxation of the solvation structure\cite{HESS_RISM} for the electronic and intramolecular structures of the solute. Contrarily, the solvation structure is adiabatically relaxed for the solute electronic structure of the solute but never for the intramolecular structure. This picture is known as the frozen approximation given by
\begin{eqnarray}
    \tilde{\mathcal{E}}({\bf R})=\braket{\Psi({\bf R})|\hat{H}_{\rm e}({\bf R})|\Psi({\bf R})}+{\bf V}({\bf R}_0){\bf d}({\bf R}) \label{eq:FRZ},
\end{eqnarray}
and it also satisfies the variational equation in Eq.(\ref{eq:gen_VP}) for ${\bf R}=0$. The difference between the relaxed and frozen approximations is expressed as follows:
\begin{eqnarray}
    \mathcal{A}({\bf R})&=&\tilde{\mathcal{E}}({\bf R})+\left[{\bf V}({\bf R})-{\bf V}(0)\right]{\bf d}({\bf R})+\mu({\bf R})-{\bf V}({\bf R}){\bf d}({\bf R})\notag\\
    &=&\tilde{\mathcal{E}}({\bf R})+\sum_{n=1}^{\infty}\frac{1}{n!}\left(\frac{\partial^n{\bf V}}{\partial {\bf R}_1\partial {\bf R}_2...\partial {\bf R}_n}\right)_{{\bf R}=0}{\bf d}({\bf R}){\bf R}_1{\bf R}_2...{\bf R}_n\notag\\
    &+&\sum_{m=2}^{\infty}\frac{1}{m!}\frac{\partial^{m}{\bf V}({\bf R})}{\partial {\bf d}_1\partial {\bf d}_2...\partial {\bf d}_{m-1}}{\bf d}({\bf R}){\bf d}_1({\bf R}){\bf d}_2({\bf R})...{\bf d}_{m-1}({\bf R})+\mathcal{F}[{\bf u}^{\rm nES},{\bm \chi};{\bf R}],
\end{eqnarray}
where $\mathcal{F}[{\bf u}^{\rm nES},{\bm \chi};{\bf R}]$ is the free energy which is a functional of nonelectrostatic potential between the solute and solvent ${\bf u}^{\rm nES}$ and the total correlation function of the solvent ${\bm \chi}$. Thus, $\mathcal{F}[{\bf u}^{\rm nES},{\bm \chi};{\bf R}]$ is independent of ${\bf V}$. Under the assumption that the deviation $\Delta {\bf V}$ is independent of ${\bf R}$ and ${\bf d}$, the derivatives of ${\bf V}$ satisfy the relation as follows:
\begin{eqnarray}
    \frac{\partial}{\partial {\bf V}}\left(\frac{\partial^n{\bf V}}{\partial {\bf R}_1\partial {\bf R}_2...\partial {\bf R}_n}\right)&=&0\\
    \frac{\partial}{\partial {\bf V}}\left(\frac{\partial^{m-1}{\bf V}}{\partial {\bf d}_1\partial {\bf d}_2...\partial {\bf d}_{m-1}}\right)&=&0.
\end{eqnarray}
Therefore, we can obtain the following easier form:
\begin{eqnarray}
    \left(\frac{\partial\mathcal{A}({\bf R})}{\partial {\bf V}}\right)_{\Delta{\bf V}=0}=\left(\frac{\partial\tilde{\mathcal{E}}({\bf R})}{\partial {\bf V}}\right)_{\Delta{\bf V}=0},
\end{eqnarray}
thus, we obtain
\begin{eqnarray}
    K_{ab,\zeta}^{(1)}=\left[\frac{\partial}{\partial V_{\zeta}}\left(\frac{\partial^2 \tilde{\mathcal{E}}}{\partial R_a\partial R_b}\right)_{{\bf R}={\bf R}_0} \right]_{\Delta{\bf V}=0}.\label{eq:K_FRZ}
\end{eqnarray}
This is also computationally easier than applying the relax approximation. Based on Eq.(\ref{eq:FRZ}), it was not necessary to calculate ${\bf h}$ and ${\bf c}$, which requires iterating self-consistent equations between the quantum chemical calculation and RISM theory by substituting the already calculated ${\bf V}({\bf R})$. By deviating ${\bf V}({\bf R}) \rightarrow {\bf V}({\bf R})+\delta {\bf V}$ to some degree and approximating Eq.(\ref{eq:K_FRZ}) by the finite-difference method, ${\bf K}^{(1)}$ can be numerically calculated as 
\begin{eqnarray}
    K_{ab,\zeta}^{(1)} \sim \frac{1}{2\delta V_{\zeta}}\left[\left\{\left(\frac{\partial^2 \tilde{\mathcal{E}}}{\partial R_a\partial R_b}\right)_{{\bf R}={\bf R}_0}\right\}_{(\Delta{\bf V})_{\zeta}=\delta V_{\zeta}}-\left\{\left(\frac{\partial^2 \tilde{\mathcal{E}}}{\partial R_a\partial R_b}\right)_{{\bf R}={\bf R}_0}\right\}_{
(\Delta{\bf V})_{\zeta}=-\delta V_{\zeta}}\right],\label{eq:K_FRZ_N}
\end{eqnarray}
using the three-point central finite difference method.\par
Notably, the given deviation $\Delta{\bf V}$ results in the adiabatically dielectric relaxation of the solute in the deployed framework. We extended the wavefunction as $\Psi({\bf R})\rightarrow \Psi({\bf R},\Delta {\bf V})$ and satisfied the extended variational equation as follows:
\begin{eqnarray}
        \Braket{\delta\Psi({\bf R},\Delta {\bf V})|\hat{H}_{e}({\bf R})+({\bf V}^t({\bf R}_0)+\Delta {\bf V}^t)\hat{\bf Q}({\bf d};{\bf R})-\hat{\Lambda}|\Psi({\bf R},\Delta {\bf V})}=0.\label{eq:gen_VP2}
\end{eqnarray}
\subsection{Spectral lines including broadening}
We assume that $\Delta{\bf V}$ are independent of the electronic and intramolecular structures of the solute and are random variables according to the Gaussian distribution\cite{Morita_RISM_Fluc,NegishiYokogawa2023} as
\begin{eqnarray}
    p(\Delta{\bf V})=\frac{1}{(2\pi)^{N/2}\sqrt{\det\left[\braket{\Delta {\bf V}\Delta{\bf V}^t}\right]}}\exp\left[-\frac{1}{2}\Delta {\bf V}^t\braket{\Delta {\bf V}\Delta{\bf V}^t}^{-1}\Delta {\bf V}\right],\label{eq:dV_gauss}
\end{eqnarray}
where $N$ is the total number of the ABSs. The matrix of the standard deviation of $\Delta {\bf V}$, $\braket{\Delta {\bf V} \Delta {\bf V}^t}$, is given by
\begin{eqnarray}
    \braket{\Delta V_\eta \Delta V_\zeta}=-k_BT\sum_{s}\rho_sq_s\int d{\bf r}\int d{\bf r}' \frac{f_{\zeta}({\bf r}')}{|{\bf r}-{\bf r}'|}\frac{\partial h_{\alpha s}(r)}{\partial d_\eta},
\end{eqnarray}
where $\zeta$ is the ABS of the atomic site $\alpha$.\par
Here, we assume that the off-diagonal terms of ${\bf K}^{(1)}$ in the representation are negligible as
\begin{eqnarray}
    K_{ij,\zeta}^{(1)}=K_{ii,\zeta}^{(1)}\delta_{ij}.
\end{eqnarray}
Thus, the spectral line $P(\omega)$ can be evaluated as follows:
\begin{eqnarray}
    P(\omega)=C\int d\Delta {\bf V} \omega^I\sum_{i}|B_i|^2\delta\left(\omega-M_i^{-\frac{1}{2}}\sqrt{K_{ii}^{(0)}+\sum_{\zeta}K_{ii,\zeta}^{(1)}\Delta V_{\zeta}}\right)p(\Delta {\bf V}) \label{eq:raw_spec}
\end{eqnarray}
where $M_i$ denotes the reduced mass of the $i$-th mode and $C$ is the prefactor. $I$ represents the frequency dependence for observing the optical processes: $I$=1, 3, and 4 corresponding to photo-absorption, photo-emission\cite{PurcellPennypacker1973,BornWolf},  and photo-scattering processes\cite{Placzek1934,BornWolf}, respectively. Furthermore, $B_i$ denotes the observables in the optical process: the first-order $i$-th mode derivative of the dipole and that of polarizability corresponding to the IR and Raman processes, respectively. In many cases, the scale of frequency fluctuation is much less than that of the eigenfrequency. Therefore, we approximate the square root part in Eq.(\ref{eq:raw_spec}) as follows:\cite{Taylor}
\begin{eqnarray}
    M_i^{-\frac{1}{2}}\sqrt{K_{ii}^{(0)}+\sum_{\zeta}K_{ii,\zeta}^{(1)}\Delta V_{\zeta}}&=&\sqrt{\omega_i^2+\sum_{\zeta}\frac{K_{ii,\zeta}^{(1)}}{M_i}\Delta V_{\zeta}}\notag\\
    &\sim&\omega_i+\sum_{\zeta}\frac{K_{ii,\zeta}^{(1)}}{2M_i\omega_i}\Delta V_{\zeta}\notag\\
    &:=&\omega_i+{\bm \kappa}_i\Delta{\bf V}.\label{eq:omg_FLC}
\end{eqnarray}
Substituting Eq.(\ref{eq:omg_FLC}) into Eq.(\ref{eq:raw_spec}) allowed us to analytically solve the approximation and the result is given as follows.
\begin{eqnarray}
    P(\omega)&=&C \omega^I\sum_{i}\frac{|B_i|^2}{({\rm FWHM})_i}\exp\left[-4\ln 2\frac{(\omega-\omega_i)^2}{({\rm FWHM})_i^2}\right] \label{eq:spec}\\
    ({\rm FWHM})_i&=&\sqrt{8\ln 2\times{\bm \kappa}_i^t\braket{\Delta{\bf V}\Delta{\bf V}^t}{\bm \kappa}_i}\label{eq:bandwidth},
\end{eqnarray}
where FWHM denotes the full-width at half-maximum of the $i$-th vibrational mode.
\section{\label{sec:level3}COMPUTATIONAL DETAILS}
We simulated acetonitrile (ACN), thio-methylnitrile(MeSCN), and benzo-nitrile(BN) molecules in various solutions using the GAMESS program package. ACN, methanol (MeOH), and water (WAT) were selected as solvents for ACN and MeSCN solutes. Acetone (ACT) and MeOH were selected as solvents for the BN solute. The CAM-B3LYP functional\cite{Yanai2004} was employed for the geometry optimization and Hessian calculations in each solvent using the RISM-SCF-cSED.\cite{Negishi_2nd,HESS_RISM} In the Hessian calculations, the analytical method was employed for all the simulations. The aug--cc--pv(D+d)Z basis sets~\cite{aug} were employed for all the atoms.\par
We applied the site--site Coulombic and Lennard--Jones(LJ) potential to
determine the intermolecular interaction between the solute and solvent using Singer and Chandler’s HNC approximation as the closure relation for the RISM. The atomic charge and LJ parameters employed for the RISM calculations were taken from OPLS-AA ~\cite{OPLSAA} and are provided with the solute and solvent parameters in Table \ref{tab: LJ_U} and \ref{tab: LJ_V}, respectively.\par
\begin{table}
\caption{\label{tab: LJ_U} Atomic charges and LJ parameters of the solutes. C(H), and C(N) denote the carbon atoms bonding with the hydrogen and nitrogen, respectively. C(CN) denotes the carbon atom that bonds with
nitrogen making nitrile of BN.}
\begin{tabular}{ccccccccc}
\hline
site& $\sigma_s$ ({\AA})& $\epsilon_s$ (kcal/mol)\\
\hline
\multicolumn{3}{c}{ACN}\\
H&3.775&0.207\\
C(H)&3.775&0.207\\
C(N)&3.650&0.150\\
N&3.070&0.170\\
\multicolumn{3}{c}{MeSCN}\\
H&3.775&0.207\\
C(H)&3.775&0.207\\
C(N)&3.650&0.150\\
N&3.070&0.170\\
S&3.550&0.250\\
\multicolumn{3}{c}{BN}\\
H&3.775&0.207\\
C(H)&3.775&0.207\\
C(CN)&3.650&0.150\\
C(N)&3.650&0.150\\
N&3.070&0.170\\
\hline
\end{tabular}
\end{table}

\begin{table}
\caption{\label{tab: LJ_V} Site charges and LJ parameters for the solvents.}
\begin{tabular}{cccc}
\hline
site& $\sigma_s$ ({\AA})& $\epsilon_s$ (kcal/mol)& $q_s$\\
\hline
\multicolumn{4}{c}{ACN}\\
CH$_3$&3.775&0.207&0.150\\
C&3.650&0.150&0.280\\
N&3.070&0.170&-0.430\\
\multicolumn{4}{c}{ACT}\\
CH$_3$&3.775&0.207&0.150\\
C&3.650&0.150&0.280\\
O&3.070&0.170&-0.430\\
\multicolumn{4}{c}{MeOH}\\
CH$_3$&3.775&0.207&0.265\\
O&3.070&0.170&-0.700\\
H&1.000&0.056&0.435\\
\multicolumn{4}{c}{WAT}\\
O&3.166&0.155&-0.820\\
H&1.000&0.056&0.410\\
\hline
\end{tabular}
\end{table}
Finally, we determined the difference $\delta {\bf V}$ to compute the numerical derivative of the Hessian. The numerical-differentiation direction was determined as the eigenvector of the matrix $\braket{\Delta {\bf V}\Delta{\bf V}^t}$. Afterward, we determined the grid size for the numerical differentiation. As the effective scale of $\Delta {\bf V}$ is given by 
\begin{eqnarray}
    \braket{|\Delta {\bf V}_{\zeta}|}^2&\sim&\braket{\Delta V_{\zeta}\Delta V_{\zeta}}\notag\\
    &=&k_BT\sum_{s}\rho_sq_s\int d{\bf r}\int d{\bf r}' \frac{f_\zeta({\bf r}')}{|{\bf r}-{\bf r'}|}\frac{\partial g_{\alpha s}(r)}{\partial d_{\zeta}},
\end{eqnarray}
we explored the scale of $\Delta{\bf V}$ roughly determined by the square root of the product of $k_BT$ and $\Delta\mu$. Employing a room temperature and solvation free energy of $k_BT\sim 10^{-3}$ a.u. and $\mu \sim 10^{-2}$ to $10^{-1}$ a.u., respectively, in many cases, the scale of $\braket{|\Delta {\bf V}_{\zeta}|}$ can be roughly estimated as $10^{-2}$ a.u. Subsequently, we applied the grid width of $\Delta V_\zeta$ as 10$^{-4}$ for the finite difference method to calculate $K_{ii,\zeta}$ to the direction of the eigenvector of $\braket{\Delta {\bf V}\Delta{\bf V}^t}$.\par
Furthermore, employing many ABSs yielded extremely small eigenvalues of $\braket{\Delta {\bf V}\Delta{\bf V}^t}$. In such cases, the numerical derivative of these eigenvectors sometimes resulted in unstabilized computations in the RISM-SCF routine. This indicates that the displacement of ${\bf V}$ to the eigenvectors is sometimes unphysical as the extremely small thermal fluctuation of ${\bf V}({\bf R})$ barely causes the displacement of ${\bf V}$. To avoid this issue, we regarded such small eigenvalues as zero when they are smaller than the threshold 10$^{-10}$ a.u. We only consider the numerical derivative of the Hessian to the direction of the eigenvector whose eigenvalue exceeds the threshold.
\section{\label{sec:level4}RESULTS AND DISCUSSION}
\subsection{Bandwidth of vibrational spectra}
\begin{figure}
    \centering
    \includegraphics[width=80mm]{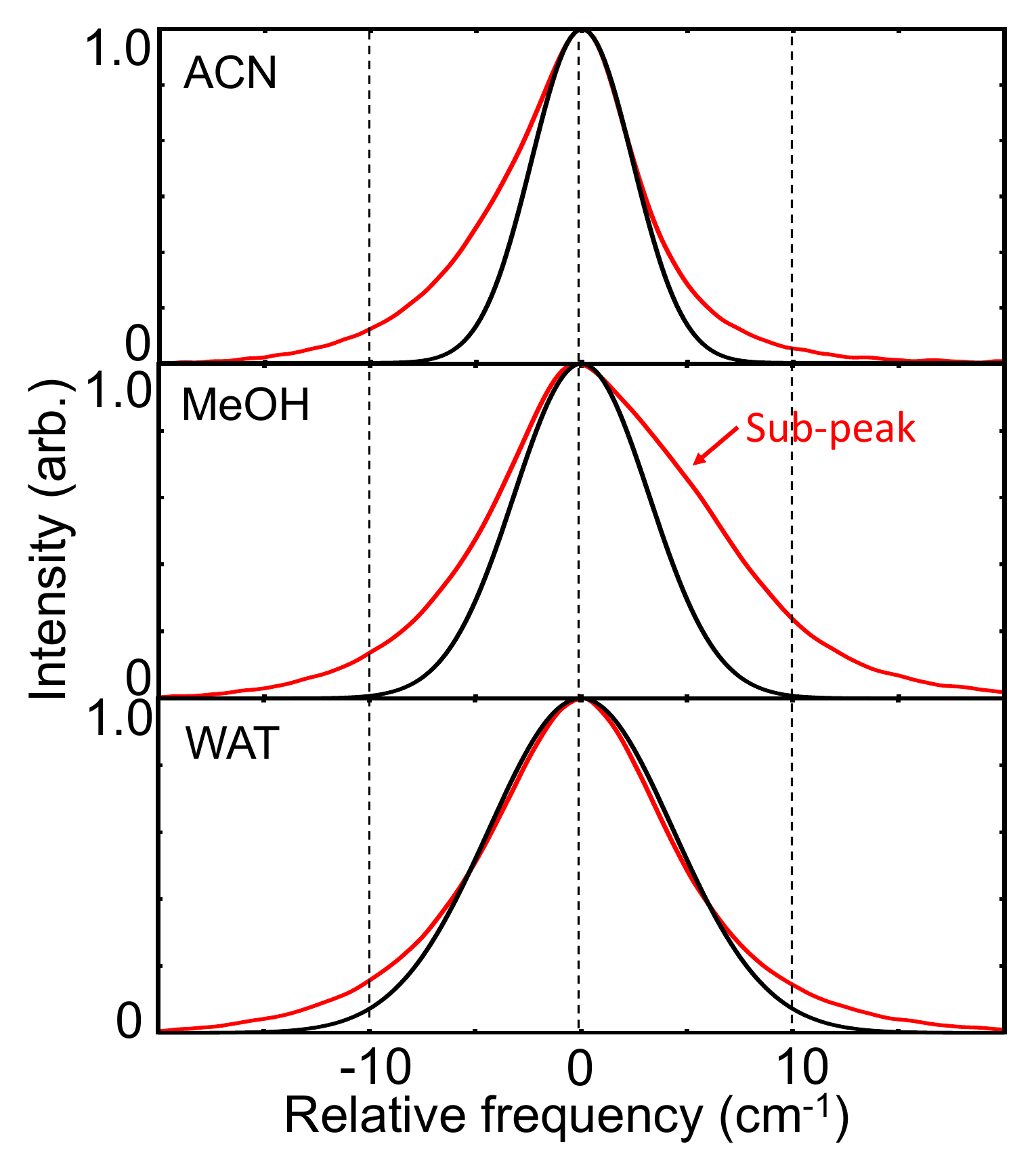}
    \caption{IR spectral lines of the C--N stretching vibration of MeSCN dissolved in various solvents. The black and red lines denote the estimation of the present theory and the experimental data, respectively.\cite{Taylor}}
    \label{fig:FIG1}
\end{figure}
First, we compare the IR spectral line of the C--N stretching of MeSCN in solvents. The spectral lines calculated using the present theory and previous experimental data\cite{EXPT2} are shown in FIG.\ref{fig:FIG1}, whose peak tops are aligned as zero frequency and peak heights are normalized as 1.0. All frequency scales of the FWHM are estimated as $\sim 10$ cm$^{-1}$. This result is consistent with the referenced experimental data. In addition, we reproduced the bandwidth-broadening trend in which the bandwidth broadened as the solvent is changed from aprotic to protic. The stochastic distribution of the electrostatic potential is assumed to be a Gaussian function; therefore, the wide tale of spectral lines in experimental data is not described in the present framework. The underestimation of the broadening in these parts indicates that the present theory did not consider other physical aspects such as the dynamics of the solvents and the other unphysical reasons depending on the experimental condition. Nevertheless, the simple assumption of the Gaussian distribution of $\Delta{\bf V}$ via simple solvation modeling such as the RISM-SCF almost surprisingly represents the experimentally observed peak-broadening scale. On the other hand, the present study failed to explain the asymmetric broadening often observed experimentally with MeOH solvent. The presence of the sub-peak is interpreted as dimerization between the solute and the solvent induced by the donation of hydrogen bonding by a proton of the MeOH. However, the interpretation of dimerization reflects the aspect of an isolated system, and there might be a gap between the interpretation and the actual system; namely, the liquid, condensed phase is more complicated. At least, representing such an asymmetric shape of the spectral line requires the extension to describe the inhomogeneous solvation to break the symmetry of the Gaussian function of Eq.(\ref{eq:dV_gauss}).\par
The present estimation and referenced experimental data\cite{EXPT1,EXPT2,EXPT3} of the bandwidth of the C-N stretching modes of the respective molecules are summarized in TABLE \ref{tab:vibACN}.
\begin{table}
\caption{\label{tab:vibACN} Eigen frequencies $\omega_i$ and FWHM of the C--N stretching vibrational modes of the three nitrile compounds dissolved in the various solvents. The experimental data are captured in circle brackets. The unit of all the quantities is cm$^{-1}$.}
\begin{tabular}{cccccc}
\hline
\multicolumn{2}{c}{ACN\footnote{The experimental data are referred by Ref.\cite{EXPT1}}}&\multicolumn{2}{c}{MeSCN\footnote{The experimental data are referred by Ref.\cite{EXPT2}}}&\multicolumn{2}{c}{BN\footnote{The experimental data are referred by Ref.\cite{EXPT3}}}\\
\hline
\hline
Solvent &FWHM&Solvent &FWHM&Solvent &FWHM\\
\hline
ACN&5.13(7.50)&ACN&5.62(7.10)&ACT&7.38(6.42)\\
MeOH&7.24(-----)\footnote{We discussed the experimental data in Ref\cite{EXCUSE}}&MeOH&7.52(-----)$^{\rm d}$&MeOH&11.6(-----)$^{\rm d}$\\
WAT&9.28(9.00)&WAT&10.2(10.0)&WAT&15.7\\
\hline
\end{tabular}
\end{table}
First, the peak-broadening scale in the present study is in 1--10 cm$^{-1}$ range for all cases and this is almost consistent with scales in the experimental data. The estimated FWHM increased as the solvents changed from aprotic to protic for all the solute molecules, as several previous experimental studies resulted in its tendency of broadening bandwidth. ACN and MeSCN solutes have been well investigated by a numerical approach using QM/MM or MD whereas BN is a complicated solute whose PES are challenging to compute and simulate. It must be noted that the present method achieved to compute such a large solute molecule whose precise computation have appeared impossible.
\subsection{Analysis of solvation fluctuation}
Here, we explored the mechanism of the peak broadening of vibrational spectra. The FWHM of the $i$-th vibrational mode quantified by Eq.(\ref{eq:bandwidth}) can be decomposed into the contributions of the solvation-field mode as follows:
\begin{eqnarray}
({\rm FWHM})_i^2&=&8\ln 2{\bm \kappa}_i^t\braket{\Delta {\bf V}\Delta {\bf V}^t}{\bm \kappa}_i\notag\\
&=&8\ln 2\sum_{\zeta\eta}{\kappa}_{i\zeta}\left[\sum_{\xi}U_{\xi\zeta}v_{\xi}U_{\xi\eta}\right]{\kappa}_{i\eta}\notag\\
    &=&8\ln 2\sum_{\xi}\sigma_{i\xi}^2\label{eq:decomp_V}\\
    \sigma_{i\xi}&:=&\sqrt{v_{\xi}}\tilde{\kappa}_{i\xi}
\end{eqnarray}
where ${\bf U}$ denotes the unitary matrix diagonalizing the matrix $\braket{\Delta {\bf V}\Delta {\bf V}^t}$; the unitary transformation is given by
\begin{eqnarray}
    \tilde{\bm \kappa}&=& {\bf U}{\bm \kappa}\\
    {\bf v}&=& {\bf U}\braket{\Delta {\bf V}\Delta {\bf V}^t}{\bf U}^{\dag}.
\end{eqnarray}
Eq.(\ref{eq:decomp_V}) allows for the decomposition of the contribution of the broadening bandwidth into the eigenmodes of solvation-field fluctuations. The FWHM of the IR spectral lines contributes to $\xi$-th fluctuation mode as $2\sqrt{2\ln 2}\times\sigma_{i\xi}$. Thus, we characterize the broadening bandwidth based on the solvation fluctuation by analysing ${\bf U}$.\par
\begin{table}
\caption{\label{tab:decomp_tab} Main and second contributions of the FWHM referred to as Eq.(\ref{eq:population}) of the C--N stretching vibrational modes of the three nitrile compounds dissolved in the various solvents.}
\begin{tabular}{cccccccccc}
\hline
Solute&\multicolumn{3}{c}{ACN}&\multicolumn{3}{c}{MeSCN}&\multicolumn{3}{c}{BN}\\
Solvent&ACN &MeOH&WAT&ACN &MeOH&WAT&ACT&MeOH&WAT\\
\hline
Main contribution&0.85&0.98&0.98&0.59&0.86&0.84&0.49&0.77&0.70\\
second contribution&0.14&0.02&0.02&0.14&0.13&0.07&0.41&0.13&0.16\\
\hline
\end{tabular}
\end{table}
We list the main and second contributions to the FWHM for each solute-solvent pair in TABLE \ref{tab:decomp_tab} as the population $p_\xi$ defined as
\begin{eqnarray}
    p_{\xi}=\frac{\sigma_{i\xi}^2}{{\bm \kappa}_i^t\braket{\Delta {\bf V}\Delta {\bf V}^t}{\bm \kappa}_i}\label{eq:population}
\end{eqnarray}
The main contributions account for over 70\% of the FWHM estimated using the present theory except for ACN-dissolved MeSCN or BN. This indicates that only one fluctuation mode can effectively describe the broadening bandwidth in the IR spectra. This result is surprising as we can simply interpret the peak broadening by only focusing on the solvation-field fluctuation of ${\bf U}$ of the main contribution mode.\par
Further, we introduce the atomic-effective solvation-field fluctuation ${\bf U}^{\rm eff}$ defined as 
\begin{eqnarray}
    U_{\xi\alpha}^{\rm eff}&:=&\frac{\sqrt{v_\xi}}{\sqrt{n_{\alpha}}}\sum_{\zeta\in\alpha}U_{\xi\zeta}\notag\\
    &=&\sqrt{v_\xi}\tilde{U}_{\xi\alpha},\label{eq:Ueff}
\end{eqnarray}
where $n_{\alpha}$ denotes the number of ABSs in $\alpha$ atomic site. This introduction is justified by the orthonormalization of ${\bf U}^{\rm eff}$. The proof is as follows:
\begin{eqnarray}
    \sum_{\xi}\tilde{U}_{\xi\gamma}\tilde{U}_{\xi\alpha}&=&\frac{1}{\sqrt{n_{\alpha}n_{\gamma}}}\sum_{\zeta\in\alpha,\eta\in\gamma}\sum_{\xi}U_{\xi\zeta}U_{\xi\eta}\notag\\
    &=&\frac{1}{\sqrt{n_{\alpha}n_{\gamma}}}\sum_{\zeta\in\alpha,\eta\in\gamma}\delta_{\zeta\eta}.\label{eq:Ueff_cond}
\end{eqnarray}
Assuming $\alpha\not=\gamma$, the summation labels of $\zeta$ and $\eta$ must not be equal so that the right-hand-side of Eq.(\ref{eq:Ueff_cond}) becomes zero. Although $\sum_{\alpha} \tilde{U}_{\xi \alpha}\tilde{U}_{\xi'\alpha}\not=\delta_{\xi\xi'}$ in general, we specifically demonstrate the orthonormalized condition based on the following assumption; $U_{\xi\zeta}=U_{\xi\eta}=u_{\xi\alpha}$ for any $\zeta$ and $\eta$ both of which are in $\alpha$ as follows:
\begin{eqnarray}
\sum_{\alpha}\tilde{U}_{\xi\alpha}\tilde{U}_{\xi'\alpha}&=&\sum_{\alpha}\sum_{\zeta\in\alpha}\sum_{\eta\in\alpha}\frac{u_{\xi\alpha}u_{\xi'\alpha}}{n_{\alpha}}\notag\\
    &=&\sum_{\alpha}u_{\xi\alpha}u_{\xi'\alpha}n_{\alpha}\notag\\
    &=&\sum_{\alpha}\sum_{\zeta\in\alpha}U_{\xi\zeta}U_{\xi'\zeta}\notag\\
    &=&\delta_{\xi\xi'}.
\end{eqnarray}
$U_{\xi\zeta}=U_{\xi\eta}=u_{\xi\alpha}$ means that the solvation-field fluctuation is independent of the shape of the electron density distribution. This assumption is almost natural because the distance between the solute and solvent significantly exceeds the scale of the existence of the solute-bound electrons.\par
\begin{figure}
    \centering
    \includegraphics[width=140mm]{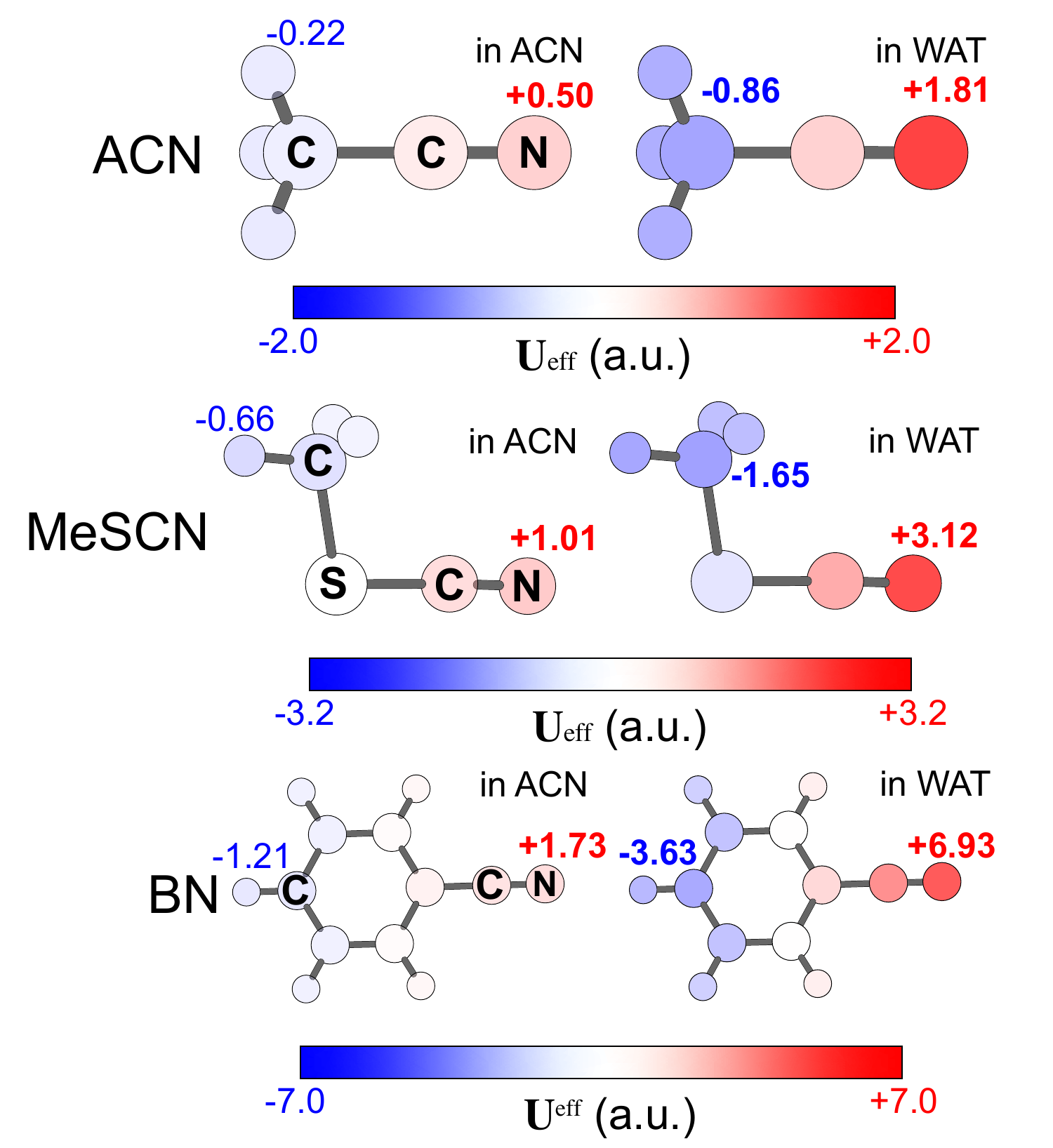}
    \caption{Main contribution mode of the electrostatic potential ${\bf U}^{\rm eff}$ for the frequency fluctuation of the C--N stretching vibration in the aprotic (ACN) and protic (WAT) solvents. The maximum and minimum values of ${\bf U}^{\rm eff}$ are represented by red and blue letters, respectively. The values in the regular and bold blue letters denote the mode amplitude of the H and C atomic sites, respectively.}
    \label{fig:FIG2}
\end{figure}
The solvation-field fluctuation of the main contribution mode ${\bf U}^{\rm eff}$ are in FIG.\ref{fig:FIG2}. The main mode exhibits a strong positive value at the N atomic site and a negative value at the opposite atomic site, e.g., the methyl site of ACN or MeSCN solute. Therefore, the main contribution of broadening bandwidth can be explained based on the fluctuation along the C--N bond axis. In addition, the protic solvents strengthen the positivity of the N atomic site than the aprotic solvents, indicating that the solvation fluctuation differs slightly in the protic and aprotic solvents. Particularly, the protic solvents promoted hydrogen bonding between their proton and the N atom of the solute, allowing for more effective local solvation fluctuation the N atom compared with the aprotic solvent.
\par
\subsection{Interpretation of vibrational-spectra broadening}
Finally, we discuss the relationship between the broadening of IR spectral lines and the main contribution mode of the solvation fluctuation (FIG.\ref{fig:FIG2}). The IR broadening is directly related to the frequency fluctuation of the CN stretching vibration. $\tilde{\kappa}_{i\xi}$ defined in Eq.(\ref{eq:omg_FLC}) is given by the first derivative of $V_{\xi}=\sum_{\zeta}U_{\xi\zeta}V_{\zeta}$ and the second derivative of the normal vibrational coordinate $Q_i$. Then, we reconstruct the effective free energy directly related to the C--N stretching vibration and solvation fluctuation characterized in FIG.\ref{fig:FIG2}.\par
Based on Eq.(\ref{eq:taylor_VR}), we only consider the harmonic oscillator of the C--N stretching mode referred to as $i$-th mode. Because the effect of thermal fluctuation of $\Delta{\bf V}$ expect the main contribution $\xi$ can be neglected, the PES of the $i$-th mode $\mathcal{A}_i({\bf R},\Delta{\bf V})$ can be approximated as follows:
\begin{eqnarray}
\mathcal{A}_i({\bf R},\Delta{\bf V})&\sim& 
    \mathcal{A}({\bf R}_0)+\frac{1}{2}K_{ii}^{(0)}Q_i^2   +A_{\xi}^{(1)}\Delta V_\xi+\tilde{G}_{i,\xi}^{(1)}Q_i\Delta V_\xi+\frac{1}{2}\tilde{K}_{ii,\xi}^{(1)}Q_i^2\Delta V_{\xi},
        \label{eq:taylor_VR2}    
\end{eqnarray}
where $\tilde{\bf G}_{\xi}^{(1)}$ and $\tilde{\bf K}_{\xi}^{(1)}$ denote the first-order differential of the gradient and the Hessian by $ V_{\xi}$ under the normal mode representation, respectively. Here, we assume that only the solvation fluctuation exists in FIG.\ref{fig:FIG2} and set the ansatz of fluctuation $\Delta {\bf V}$ as one dimensional voltage $\Delta V(x)$ using the constants $L$ and $\epsilon$ as follows:
\begin{eqnarray}
    \Delta V(x)=-\frac{\Delta V_{\xi}}{L}x(1+\epsilon x)+O(x^2),
\end{eqnarray}
where $x$ denotes the one-dimensional spatial coordinate along the C--N bond axis and $x=0$ is defined as the center of the permanent dipole of the solute. Further, $L$ denotes the distance between the sites of the lowest and highest voltages. The fluctuation of the one-dimensional voltage $\Delta V$ is proportional to $\Delta V_\xi$. Regarding the $\alpha$-th atomic site charges $q_\alpha(Q_i)$ as a function of $Q_i$, the fluctuation of the excess energy $\Delta \mu_{\rm ex}(Q_i)$ corrected by the solvation effect is given by
\begin{eqnarray}
   \Delta \mu_{\rm ex}(Q_i)=\sum_{\alpha}q_\alpha(Q_i)\Delta V[x_\alpha(Q_i)].\label{eq:Deltaq}
\end{eqnarray}
where $x_\alpha$ is the position of $\alpha$ site and linearly depends on $Q_i$. As the enthalpy of the electrostatic contribution is given by Eq.(\ref{eq:Deltaq}) while the fluctuation of entropy is neglected in the system, the fluctuation of the PES corresponds to the fluctuation of the excess energy as follows:
\begin{eqnarray}
    \frac{\partial^3\mathcal{A}_i}{\partial Q_i^2\partial V_{\xi}}=\frac{\partial^3\Delta \mu_{\rm ex}}{\partial Q_i^2\partial V_{\xi}}.
\end{eqnarray}
Therefore, the above equation yields
\begin{eqnarray}
    \tilde{K}_{ii,\xi}^{(1)}\Delta{V}_{\xi}&=&\Delta{V}_{\xi}\left[\frac{\partial^3\Delta \mu_{\rm ex}}{\partial Q_i^2\partial V_{\xi}}\right]_{Q_i=0,\Delta {\bf V}=0}\notag\\
    &=&\frac{\Delta{V}_{\xi}}{L}\sum_{\alpha}\left\{\epsilon q_{\alpha}(0)+\left(\frac{dq_{\alpha}}{dQ_i}\right)_{Q_i=0}\right\}\left(\frac{dx_{\alpha}}{dQ_i}\right)_{Q_i=0}+O(Q_i).\label{eq:minimal_eff}
\end{eqnarray}
The first and second terms of Eq.(\ref{eq:minimal_eff}) denote the solvation-induced electric field non-uniformity and the bond-length-change-induced effect of the electron transfer. Owing to the generally minimal effect of charge transfer on the vibrational spectra, we neglected the second term in Eq.(\ref{eq:minimal_eff}).\par
The non-uniformity of the electric field stems from the locality of the solvation structure. In particular, protic solvents form microsolvations such as hydrogen bonding and generate a strong electrostatic fields around the N atom, an acceptor of the proton. This corresponds to their larger second-order coefficient $\epsilon$ compared to those of aprotic solvents. As noted in the solvation fluctuation in FIG.\ref{fig:FIG2}, WAT exhibits a larger amplitude at the N atomic site than ACN. In summary, an electric field generated by hydrogen bonding is adjusted as the potential energy of a quadratic function, significantly influencing the changes in the intramolecular vibrational frequencies. Thus, we conclude that the fluctuations of the electric field intensity
$E$ induced by the solvation effect are directly connected to the vibrational frequency fluctuations.
\section{\label{sec:level5}CONCLUSION}
By extending the electrostatic potential to random variables based on the RISM-SCF-cSED framework, we formulated the frequency fluctuations of intramolecular vibrations of solutes. Based on this framework, we evaluate a part of the broadening of vibrational spectral lines of various solution systems. Further, we applied the present theory to C--N stretching vibration of several nitrile compounds dissolved in polar solvents. The estimated frequency-fluctuation scales are consistent with those in the referenced experimental data. The protic solvents broaden the IR absorption bandwidth via hydrogen bonding in the present study. The present method exhibits great advantages regarding the computational cost. It easily computes benzonitrile comprising 13 atomics sites whereas other numerical frameworks only reported the simulation results of small-sized solute molecules such as MeSCN.\par
The present theory enables allowed for the analysis of the relationship between the frequency and thermal fluctuations of solutes and solvation fields, respectively. We presented a scheme for evaluating the matrices of the standard deviation of the electrostatic potential regarded as the multiple random variables. Diagonalization of the matrices to decompose the frequency fluctuation into the contributions of solvation fluctuation clarified that only one solvation-fluctuation mode dominated the frequency fluctuation in the case of C--N stretching vibrations. The mode is almost parallel to the vibrational axis. In particular, as the solvents form stronger hydrogen bonds with the solutes, the electrostatic potential fluctuations became spatially non-uniform. Thus, we assumed that the solvation-induced energy correction is characterized by the coupling energy between the permanent dipole of C--N stretching vibration and the electrostatic field generated by the solvents. Overall, we conclude that the coupling non-uniformity of the electrostatic field and the strength of the permanent dipole were crucial to the characterization of frequency fluctuation.
%今回のこの解釈が他の分子系でも成立するのであれば、極性溶媒中における振動スペクトルのブロードニングと振動方向の電場揺らぎの空間的な非一様性には強い相関があることを論じることができる。これは、ミクロ溶媒和と分子振動状態のカップリングに関する新しい知見を与えうるという意味で意義深いものとなるだろう。
\begin{acknowledgments}
This study was supported by MEXT KAKENHI Grants (Green Catalysis Science) JP23H04911 (D.Y.) and Grant--in--Aid for JSPS Research Fellow No. 22KJ1094 (N.N.). The authors would like to
thank Enago (www.enago.jp) for the English review.
\end{acknowledgments}
\section*{\bf DATA ABAILABILITY}
The data that support the findings of this study are available from the corresponding author upon reasonable request.
\providecommand{\noopsort}[1]{}\providecommand{\singleletter}[1]{#1}%\providecommand{\noopsort}[1]{}\providecommand{\singleletter}[1]{#1}%


\begin{thebibliography}{41}
\bibitem [{EXC()}]{EXCUSE}%
  \BibitemOpen
  \href@noop {} {}\bibinfo {note} {As we mentioned in this subsection, several
  previous experiments have shown the double peak or shoulder structure in the
  spectral lines. The detail of the peaks structure is an open question and
  thus the data of FWHM is out of the range of the present
  discussion.}\BibitemShut {Stop}%
\end{thebibliography}
\end{document}